\documentclass[letterpaper, 10 pt, conference]{ieeeconf}  % Comment this line out
\IEEEoverridecommandlockouts                              % This command is only
\title{\LARGE \bf
Maximum A Posteriori Learning in Demand Competition Games
}

\author{Mohsen Rakhshan}

\usepackage{multirow}
\usepackage{latexsym}
\usepackage{amsmath}
\usepackage{amsfonts}
\usepackage{amssymb}
\usepackage{graphicx}
\usepackage{times}
\usepackage{cite}

\begin{document}

\maketitle
\thispagestyle{empty}
\pagestyle{empty}

%%%%%%%%%%%%%%%%%%%%%%%%%%%%%%%%%%%%%%%%%%%%%%%%%%%%%%%%%%%%%%%%%%%%%%%%%%%%%%%%
\begin{abstract}
We consider an inventory competition game between two firms. The question we address is this: If players do not know the opponent's action and opponent's utility function can they learn to play the Nash policy in a repeated game by observing their own sales? In this work it is proven that by means of Maximum A Posteriori (MAP) estimation, players can learn the Nash policy. It is proven that players' actions and beliefs do converge to the Nash equilibrium.
\end{abstract}

%%%%%%%%%%%%%%%%%%%%%%%%%%%%%%%%%%%%%%%%%%%%%%%%%%%%%%%%%%%%%%%%%%%%%%%%%%%%%%%%
\section{INTRODUCTION}

Assume two players are engaged in a strategic learning process in which they play a one-stage game repeatedly. In a one-stage game, at the beginning of every stage players order inventory levels by incurring ordering costs, and then a random demand occurs. If a player meets the demand she/he collects revenues. If a player has extra inventory levels at the end of the stage, she/he will incur holding costs. A proportion of any unmet demand will switch to another player. In this game the objective of every player is to make ordering decisions so as to maximize her/his own expected revenue.

We consider a scenario in which every player is informed of her/his own utility function but does not know the opponent's utility function. Each player knows both her/his own local demand distribution and the opponent's local demand distribution, but she/he cannot observe the current and past actions of the opponent. Players observe their own sales and remember their own previous sales. Players' sales contain information about the opponent's action. Each player constructs a belief about the opponent's strategy set such that includes opponent's Nash policy; i.e., player $i$ ($P_i$) assumes that the opponent plays a threshold policy with a uniform distribution over $[0,a_j]$. A player's belief is a conditional probability density function over the opponent's strategies given the previous sales. At every stage of the repeated game, players observe their own sales and then update their beliefs about their opponent's strategy set. At every stage of the repeated game, every player has a belief about the opponent's strategy set. At every stage of the repeated game, players compute the Maximum A Posteriori (MAP) estimation of their belief and play their best response to that strategy.

The studies most related to this work are \cite{Z1}-\cite{Z4}. Authors in $[2],[3],$ and $[4]$ address the competitive inventory game. %According to the authors knowledge, this is the first work on the learning of Nash equilibrium in a competitive inventory game.
Various authors have argued the convergence to Nash equilibrium in games with a finite strategy set (see $[5],[6]$ and $[7]$). %In this work the strategy set is infinite; this situation is considerably more complex than the case with a finite strategy set.
{\subsection{Notation}
\begin{eqnarray*}
&&(.)^+=\max(.,0).\\
&&d_{i}: \text{Local demand faced by player} \,\ i=1,2,\\
&&\text{are non-negative-valued independent random variables},\\
&&\text{with a continuous density function}.\\
&&f_{d_i}: \text{Local demand's density function for} \,\ P_i; \,\ i=1,2.\\
&&\xi_i^n: \text{Random outcome for local demand} \,\ d_i \,\ \text{at stage} \,\ n. \\
&&c_i: \text{Unit variable ordering cost for} \,\ P_i\,\ (c_i>0).\\
&&h_i: \text{Unit holding cost for} \,\ P_i \,\ (h_i>0).\\
&&r_i: \text{Unit selling price for} \,\ P_i \,\ (r_i>0).\\
&&y_i: P_i \text{'s action in one-stage game} \,\ (y_i \geq 0).\\
&&\alpha_i: \text{Proportion of unmet demand that switches from} \\
&&P_j \,\ \text{to} \,\ P_i.\\
&&\bar{d}_{i}:=d_i+\alpha_i \,\ (d_j-y_j)^+, \,\ P_i\text{'s total demand}.\\
&&F_{\bar{d}_i(y_j)}: \text{Cumulative distribution function of}\,\ P_i\text{'s total}\\
&&\text{demand given} \,\ y_{j}.\\
&&g_i(y_1,y_2):=r_i\min \{y_i,\bar d_i\}-h_i (y_i-\bar d_i)^+-c_iy_i.\\
&&G_i(y_1,y_2):=E_{d_1,d_2} \,\ g_i(y_1,y_2), P_i\text{'s utility function where}\\
&&E_{d_1,d_2} \,\ \text{means expectation over local demands} \,\ d_1 \,\ \text{and} \,\ d_2.\\
&&y_{i}^n: \text{Optimal inventory level for} \,\ P_i \,\ \text{at stage n, where} \\
&&y_i^n=\arg\max_{y_i} G_i(y_i,\bar{y}_j), \text{in which}\\
&&\bar{y}_j=\arg \max_{y_j} f^n(y_j|I_i^n).\\
&&s_i^n: P_i \text{'s sales at stage} \,\ n,\\
&&s_i^n=\min(y_i^n,\xi_i^n+(\xi_j^n-y_j^n)^+),\,\ i \neq j=1,2.\\
&&I_i^n: P_i \text{'s information vector at stage} \,\ n, I_i^1=\big \{g_i(.)\big \},\\
&&I_i^{n+1}=I_i^n \cup \big \{ (s_i^n,y_i^n) \big \}, n=1,2,....\\
&&BR_i(.): P_i \text{'s best response to '.', where}\\
&&BR_i(.)=\arg\max_{y_i}G_i(y_i,.).
\end{eqnarray*}

\subsection{One-Stage Game}
Consider two players $i\in\{1,2\}$ without initial inventory
levels. Each player $i$ brings his inventory level to $y_i$, where $y_i\geq 0$, by incurring the total ordering cost $c_iy_i$, where $c_i\geq0$ is the variable ordering cost per unit. Then each player $i$ receives the random
demand $d_i$ from the customers for whom player $i$ is the first
choice. If the demand $d_j$ is unmet by player $j$, i.e., $d_j>
y_j$, then a fixed proportion $\alpha_{i}\in[0,1]$ of this unmet
demand $(d_j-y_j)^+$ switches to player $i \neq j$. Therefore, the total
demand faced by player $i$ becomes
$$\bar{d}_i(y_{j})=d_i+\alpha_{i}(d_{j}-y_{j})^+$$
in which $(.)^+=max(.,0)$. We will suppress the dependence of $\bar{d}_i$ on $y_{j}$ when
there is no possibility of confusion. Subsequently, player $i$ collects the revenues $r_i\min\{y_i,\bar{d}_i\}$ and incurs the holding cost $h_i(y_i-\bar{d}_i)^+$ where $r_i>0$ is the unit
revenue and $h_i>0$ is the unit holding cost. As a result, player
$i$'s utility function is given as
$$G_i(y_1,y_2)=E_{d_1,d_2} \,\ g_i(y_1,y_2),$$
where the expectation is taken with respect to $(d_1,d_2)$ and
$$g_i(y_1,y_2)=\\
r_i\min \{y_i,\bar d_i\}-h_i (y_i-\bar d_i)^+-c_iy_i.$$
We denote the game characterized by the utility functions given above and the strategy
sets $\{[0,\infty)\}_{i\in\{1,2\}}$ as $\Gamma$.

\subsection{Repeated Games}

Player's utility function, given actions $(y_1,y_2)$, is
\begin{eqnarray*}
G_1(y_1,y_2)=E_{d_1,d_2} \,\ g_1(y_1,y_2),&&\\
G_2(y_1,y_2)=E_{d_1,d_2} \,\ g_2(y_1,y_2);&&
\end{eqnarray*}
the expectation is over local demands $d_1$ and $d_2$, in which
\begin{eqnarray*}
\bar d_1=d_1+\alpha_1 (d_2-y_2)^+,&&\\
\bar d_2=d_2+\alpha_2 (d_1-y_1)^+,&&
\end{eqnarray*}
\begin{eqnarray*}
g_1(y_1,y_2)=r_1\min \{y_1,\bar d_1\}-h_1 (y_1-\bar d_1)^+-c_1 y_1,&&\\
g_2(y_1,y_2)=r_2\min \{y_2,\bar d_2\}-h_2 (y_2-\bar d_2)^+-c_2 y_2.&&
\end{eqnarray*}
In repeated games, players try to maximize their own utility function given their belief about their opponent's strategy. Assume the information vectors $\{I_1^n,I_2^n\}$ and beliefs $\{f^n(y_1|I_2^n),f^n(y_2|I_1^n)\}$ at stage $n$ are given. The learning process at stage $n$ is given below:
\begin{enumerate}
\item Players choose optimal actions $(y_1^n,y_2^n)$, given their belief about the opponent's action:
\begin{eqnarray*}
\bar{y}_1&=&\arg \max_{y_1} f^n(y_1|I_2^n),\\
\bar{y}_2&=&\arg \max_{y_2} f^n(y_2|I_1^n),\\
y_1^n&=&\arg\max_{y_1} G_1(y_1,\bar{y}_2), \\
y_2^n&=&\arg\max_{y_2} G_2(\bar{y}_1,y_2).
\end{eqnarray*}

\item Demands occur, and players observe their own sales:
\begin{eqnarray*}
&&s_1^n=\min(y_1^n,\xi_1^n+(\xi_2^n-y_2^n)^+), \\
&&s_2^n=\min(y_2^n,\xi_2^n+(\xi_1^n-y_1^n)^+)
\end{eqnarray*}
where $\xi_1^n$ and  $\xi_2^n$ are defined as random outcomes respectively for local demands $d_1$ and $d_2$ at stage $n$ with density functions $f_{d_1}$ and $f_{d_2}$.
\item Players update their information vectors:
\begin{eqnarray*}
I_1^{n+1}&=&I_1^{n} \cup \big \{ (s_1^{n},y_1^{n})  \big \},\\
I_2^{n+1}&=&I_2^{n} \cup \big \{ (s_2^{n},y_2^{n})  \big \}.
\end{eqnarray*}

\item Players update their beliefs about the distribution of the opponent's strategy:
\\

$f^{n+1}(y_1|I_2^{n+1})=\frac{f(s_2^n|y_1,y_2^n)f^n(y_1|I_2^{n})}{\int f(s_2^n|y_1,y_2^n)f^n(y_1|I_2^{n})d y_1},$\\
\\
$f^{n+1}(y_2|I_1^{n+1})=\frac{f(s_1^n|y_2,y_1^n)f^n(y_2|I_1^{n})}{\int f(s_1^n|y_2,y_1^n)f^n(y_2|I_1^{n})d y_2}.(1)$  \\
\end{enumerate}
The learning process at stage $n+1$ is similar to stage $n$.
In this work it will be shown that $(y_1^n,y_2^n) \rightarrow (y_1^*,y_2^*)$ where $(y_1^*,y_2^*)$ is the Nash equilibrium:
\begin{eqnarray*}
y_1^*=\arg\max_{y_1}\,\ G_1(y_1,y_2^*)=\arg\max_{y_{1}} \,\ E \,\ g_1(y_1,y_2^*),&&\\
y_2^*=\arg\max_{y_2}\,\ G_2(y_1^*,y_2)=\arg\max_{y_2} \,\ E \,\ g_2(y_1^*,y_2),&&
\end{eqnarray*}
in which
\begin{eqnarray*}
\bar d_1^*=d_1+\alpha_1(d_2-y_2^*)^+,&&\\
\bar d_2^*=d_2+\alpha_2(d_1-y_1^*)^+,&&
\end{eqnarray*}
\begin{eqnarray*}
g_1(y_1,y_2^*)=r_1\min \{y_1,\bar d_1^*\}-h_1 (y_1-\bar d_1^*)^+-c_1 y_1,&&\\
g_2(y_1^*,y_2)= r_2\min \{y_2,\bar d_2^*\}-h_2 (y_2-\bar d_2^*)^+-c_2 y_2.&&
\end{eqnarray*}

%%%%%%%%%%%%%%%%%%%%%%%%%%%%%%%%%%%%%%%%%%%%%%%%%%%%%%%%%%%%%%%%%%%%%%%%%%%%%%%%

\section{Discrete Learning Model}

At the first stage there is no sales history, and the information vector is given as $I_i^1=\big \{g_i(.) \big \}$. $P_i$ supposes that $P_j$ plays a threshold policy with a uniform distribution on $[0,a_j]$, therefore the initial belief is the continuous density function $f^1(y_j|I_i^1)=1/a_j$ for $0 \leq y_j \leq a_j$.
It is noticeable that $y_{j}^{*}\in[0,a_j]$. To make the model discrete, for example, let step size be $\Delta$. $P_i$'s discrete belief about the distribution of $P_j$'s strategy at stage $n$ is $\mu_i^n=[\mu_i^n(1)\,\ \mu_i^n(2)\cdots \mu_i^n(M_i)]$, where $M_i=\lceil\frac{a_j}{\Delta}\rceil$ and $\mu_i^n(k)$, $k=1,...,M_i$ stands for the probability that $P_j$'s strategy will be $(k-1)\Delta \leq y_j < k\Delta$, given $I_i^n$. It is noticeable that $\sum_{j=1}^{M_i}\mu_{i}^n(j)=1$.

At the first stage, $P_i$'s initial discrete belief about $P_j$'s action, $y_j$, is $\mu_i^1=[\mu_i^1(1) \,\ \mu_i^1(2) \cdots \mu_i^1(M_i)]$, in which $$\mu_i^1(k)=\frac{\int_{(k-1)\Delta}^{k\Delta} f^1(y_j|I_i^1)dy_j}{\Delta}.$$
At the first stage, players choose the optimal actions given their belief about the distribution of the opponent's strategy. Then demands occur, and players observe their own sales and update their belief about the distribution of the opponent's strategy based on Bayes's rule.

At the second stage, the information vector is $I_i^2=\big \{I_i^1,(s_i^1,y_i^1) \big \}$, and $P_i$'s discrete belief about opponent's action is $\mu_i^2=[\mu_i^2(1) \,\ \mu_i^2(2) \cdots \mu_i^2(M_i)]$, where \\

$\mu_i^2(k)=\frac{\int_{(k-1)\Delta}^{k\Delta}f^2(y_j|I_i^2)dy_j}{\Delta},\,\
    k=1,2,...,M_i.  \,\ (2)$\\
\\
From (1),\\

$f^2(y_j|I_i^2)=\frac{f(s_i^1|y_j,y_i^1)f^1(y_j|I_i^1)}{\int_{0}^{a_j} f(s_i^1|y_j,y_i^1)f^1(y_j|I_i^1)dy_j}.    \,\ \,\ (3)$\\
\\
From (2) and (3),\\

$\mu_i^2(k)=\frac{\int_{(k-1)\Delta}^{k\Delta}f(s_i^1|y_j,y_i^1)f^1(y_j|I_i^1)dy_j}{\Delta\int_{0}^{a_j} f(s_i^1|y_j,y_i^1)f^1(y_j|I_i^1)dy_j}$, \\

$\mu_i^2(k) \approx \frac{\mu_{i}^{1}(k)\int_{(k-1)\Delta}^{k\Delta}f(s_i^1|y_j,y_i^1)dy_j}{\Delta\sum_{l=1}^{M}\mu_i^1(l)\int_{(l-1)\Delta}^{l\Delta}f(s_i^1|y_j,y_i^1)dy_j}$.
\\
\\
Discrete local demands are\\

$P_{d_1}(k)=\frac{\int_{(k-1)\Delta}^{k\Delta}f_{d_1}d d_1}{\Delta}$ for $k=1,...,N_1$,\\

$P_{d_2}(k)=\frac{\int_{(k-1)\Delta}^{k\Delta}f_{d_2} d d_2}{\Delta}$ for $k=1,...,N_2$.\\
\\
It is noticeable that \\

$\sum_{k=1}^{N_1}P_{d_1}(k)=1$, \\

$\sum_{k=1}^{N_2}P_{d_2}(k)=1$, \\

$\int_{(k-1)\Delta}^{k\Delta}f(s_i^1|y_j,y_i^1)dy_j=\sum_{(l,m) \in A} \frac{P_{d_1}(l) P_{d_2}(m)}{\Delta^2}$, \\
\\
where $A$ is defined as\\

$A=\{(l,m) \big{|} \,\  1 \leq l \leq N_1, \,\  1 \leq m \leq N_2 \,\ s_i^1 \in [ a ,b ]\}$,\\

$a=min\big(y_i^1,(l-1)\Delta+\alpha_i ((m-1)\Delta-k\Delta)^+\big)$,\\

$b=min\big(y_i^1,l\Delta+\alpha_i (m\Delta-(k-1)\Delta)^+\big)$.\\

\textbf{Example 1}: Consider the case $r_i=4$, $h_i=0.6$, $\alpha_i=1$, $c_1=2$, $c_2=1$ and $f(d_i)=1$ for $0 \leq d_i \leq 1$, $i=1,2$. To make
the model discrete, let step size be $\Delta=0.01$. In figures 1 and 2, it is shown that by means of MAP estimation players' actions and beliefs converge to the Nash equilibrium $(0.45,0.79)$.

\begin{figure}[thpb]
      \centering
      \includegraphics[width=3in]{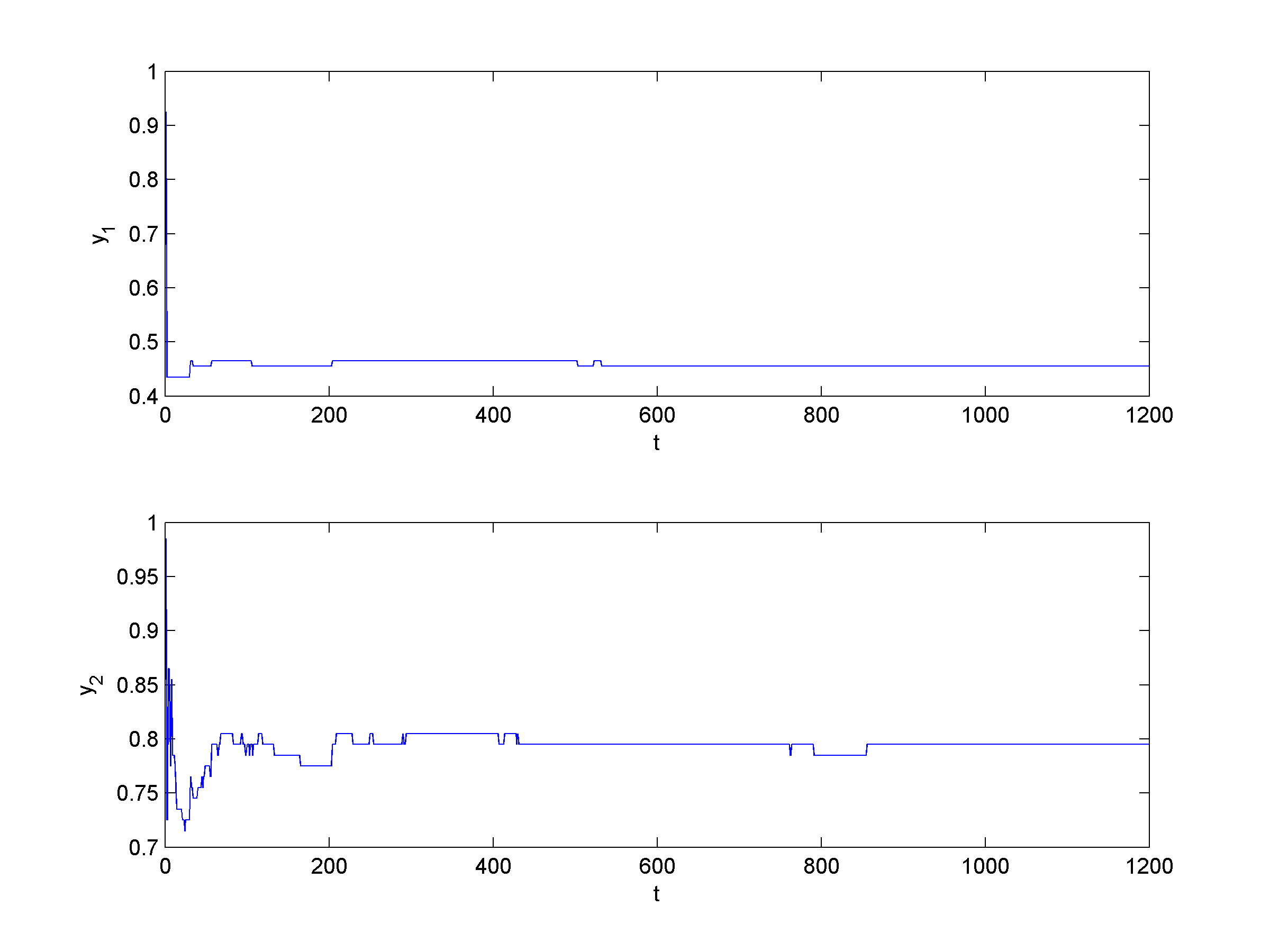}
      \caption{Convergence of actions to the Nash equilibrium $(0.45,0.79)$}
      \label{figurelabel}
\end{figure}

\begin{figure}[thpb]
      \centering
      \includegraphics[width=3in]{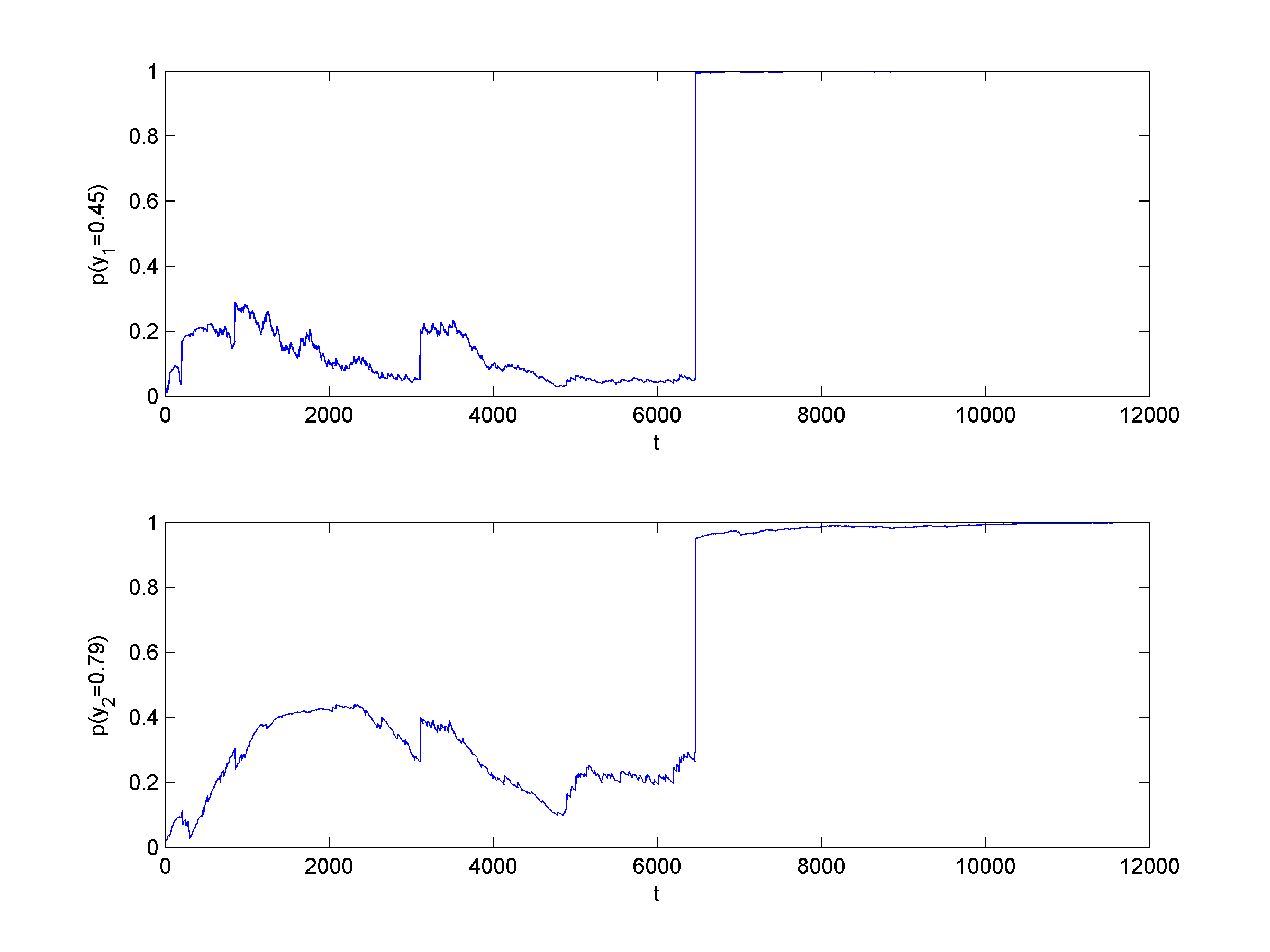}
      \caption{Convergence of beliefs to the Nash equilibrium $(0.45,0.79)$}
      \label{figurelabel}
\end{figure}

\section{Convergence to the Nash Equilibrium}

At the first stage, $P_1$ believes that $P_2$ will choose a policy in the interval $B_1=[\underline{b}_1,\overline{b}_1]$ with density function $f^1(y_2|I_1^1)$, and similarly, $P_2$ believes that $P_1$ will choose a policy in the interval $B_2=[\underline{b}_2,\overline{b}_2]$ with density function $f^1(y_1|I_2^1)$. Proposition 1 means that there exists an interval $U \subset B_1$ such that $P_2$ will not play any policy in $U$.

\textbf{Proposition 1}: For at least one of the players, without loss of generality for $P_1$, there exists an interval $U=[\underline{b}_1 \,\ \underline{\beta})$ or $U=(\overline{\beta} \,\ \overline{b}_1]$ or $U=[\underline{b}_1 \,\ \underline{\beta}) \cup (\overline{\beta} \,\ \overline{b}_1]$, such that $y_2^n \notin U \subset B_1=[\underline{b}_1 \,\ \overline{b}_1]$ for $n=1,2,...$. This means that $P_2$ will never play any policy in $U$. Proof is given in the Appendix.

It is noticeable that the set $U$ can always be found unless $\underline{b}_1=\overline{b}_1=y_2^*$ and $\underline{b}_2=\overline{b}_2=y_1^*$.

Proposition 2 means that if there exists an interval $U \subset B_1$ such that $P_2$ does not choose any policy in $U$ then $P_1$ will eventually figure out that $y_2^n \notin U$.

\textbf{Proposition 2}: If there exists an interval $U$, without loss of generality $U=(\overline{\beta} \,\ \overline{b}_1]$, such that $y_2^n \notin U$ for $n=N,N+1,...$, then for any given $\epsilon > 0$,
\begin{eqnarray*}
P\big(\lim_{n\longrightarrow \infty} \,\ \arg\max_{y_2} \,\ f^n(y_2|I_1^n)\notin (\bar{\beta}+\epsilon \,\ \overline{b}_1]\big)=1.
\end{eqnarray*}
Proof is given in the Appendix. \\

\textbf{Theorem}: Players' actions in the demand competition game will converge to the Nash equilibrium, $(y_1^*,y_2^*)$, by MAP.

Proof: At the first stage, $P_1$ believes that $P_2$ will choose a policy in the interval $B_1=[\underline{b}_1^1,\overline{b}_1^1]$ with density function $f^1(y_2|I_1^1)$, and similarly, $P_2$ believes that $P_1$ will choose a policy in the interval $B_2=[\underline{b}_2^1,\overline{b}_2^1]$ with density function $f^1(y_1|I_2^1)$. \\
Without loss of generality, assume
\begin{eqnarray*}
BR_2(\underline{b}_2^1)<\overline{b}_1^1,\\
\underline{b}_2^1<BR_1(\overline{b}_1^1).
\end{eqnarray*}
Here the worst-case is considered, in which
\begin{eqnarray*}
y_1^*=\overline{b}_2^1,\\
y_2^*=\underline{b}_1^1.
\end{eqnarray*}
According to Proposition 2, for the given $0 < \epsilon_1 < \overline{b}_1^1-BR_2(\underline{b}_2^1)$ there exists an $N_1$ such that $P(A_{N_1,\epsilon_1})=1$,
\begin{eqnarray*}
A_{n,\epsilon}=\{ \arg\max_{y_2} \,\ f^k(y_2|I_1^k) \in [\underline{b}_1^1 \,\ BR_2(\underline{b}_2^1)+\epsilon] | \forall k \geq n\}.
\end{eqnarray*}
According to Proposition 2, for the given $0 < \epsilon_2 < BR_1(\overline{b}_1^1)-\underline{b}_2^1$ there exists an $N_2$ such that $P(B_{N_2,\epsilon_2})=1,$
\begin{eqnarray*}
B_{n,\epsilon}=\{ \arg \max_{y_1} \,\ f^k(y_1|I_2^k) \in [BR_1(\overline{b}_1^1)-\epsilon \,\ \overline{b}_2^1] | \forall k \geq n \}.
\end{eqnarray*}
Let define,
\begin{eqnarray*}
&&[\underline{b}_1^2 \,\ \overline{b}_1^2] := [\underline{b}_1^1 \,\ BR_2(\underline{b}_2^1)+\epsilon_1],\\
&&[\underline{b}_2^2 \,\ \overline{b}_2^2] := [BR_1(\overline{b}_1^1)-\epsilon_2 \,\ \overline{b}_2^1],\\
&&n_1=\max(N_1,N_2).
\end{eqnarray*}
According to Proposition 2, for $n \geq n_1$, $P(A_{n,\epsilon_1})=1$ and \\
$P(B_{n,\epsilon_2})=1$; therefore,
\begin{eqnarray*}
\arg \max_{y_2} \,\ f^k(y_2|I_1^k) \in [\underline{b}_1^2 \,\ \overline{b}_1^2] \,\  \forall \,\ k \geq n_1,\\
\arg \max_{y_1} \,\ f^k(y_1|I_2^k) \in [\underline{b}_2^2 \,\ \overline{b}_2^2] \,\  \forall \,\ k \geq n_1.
\end{eqnarray*}
It is evident that $\overline{b}_1^2 < \overline{b}_1^1$ and $\underline{b}_2^2 > \underline{b}_2^1$. According to Proposition 1, without loss of generality, assume
\begin{eqnarray*}
BR_2(\underline{b}_2^2) < \overline{b}_1^2,\\
\underline{b}_2^2 < BR_1(\overline{b}_1^2).
\end{eqnarray*}
According to Proposition 2, for the given $0 < \epsilon_3 < BR_2(\underline{b}_2^1)-BR_2(\underline{b}_2^2)$ there exists an $N_3$ such that $P(A_{N_3,\epsilon_3})=1$,
\begin{eqnarray*}
A_{n,\epsilon}=\{ \arg \max_{y_2} \,\ f^k(y_2|I_1^k) \in [\underline{b}_1^2 \,\ BR_2(\underline{b}_2^2)+\epsilon]| \forall k \geq n \}.
\end{eqnarray*}
According to Proposition 2, for the given $0 < \epsilon_4 < BR_1(\overline{b}_1^2)-BR_1(\overline{b}_1^1)$ there exists an $N_4$ such that $P(B_{N_4,\epsilon_4})=1$,
\begin{eqnarray*}
B_{n,\epsilon}=\{ \arg \max_{y_1} \,\ f^k(y_1|I_2^k) \in [BR_1(\overline{b}_1^2)-\epsilon \,\ \overline{b}_2^2]| \forall k \geq n \}.
\end{eqnarray*}
Let define,
\begin{eqnarray*}
&&[\underline{b}_1^3 \,\ \overline{b}_1^3] := [\underline{b}_1^2 \,\ BR_2(\underline{b}_2^2)+\epsilon_3],\\
&&[\underline{b}_2^3 \,\ \overline{b}_2^3] := [BR_1(\overline{b}_1^2)-\epsilon_4 \,\ \overline{b}_2^2],\\
&&n_2=\max(N_3,N_4).
\end{eqnarray*}
According to Proposition 2, for $n \geq n_2$, $P(A_{n,\epsilon_3})=1$ and \\
$P(B_{n,\epsilon_4})=1$; therefore,
\begin{eqnarray*}
&&\arg \max_{y_2} \,\ f^k(y_2|I_1^k) \in [\underline{b}_1^3 \,\ \overline{b}_1^3] \,\  \text{for} \,\ \forall k \geq n_2,\\
&&\arg \max_{y_1} \,\ f^k(y_1|I_2^k) \in [\underline{b}_2^3 \,\ \overline{b}_2^3] \,\ \text{for} \,\ \forall k \geq n_2.
\end{eqnarray*}
According to Proposition 1, without loss of generality, assume
\begin{eqnarray*}
BR_2(\underline{b}_2^3) < \overline{b}_1^3,\\
\underline{b}_2^3 < BR_1(\overline{b}_1^3).
\end{eqnarray*}
Because $\underline{b}_2^3 > BR_1(\overline{b}_1^1)$ and because according to Proposition 2, for the given $0 < \epsilon_5 < BR_2(BR_1(\overline{b}_1^1))-BR_2(\underline{b}_2^3)$ there exists an $N_5$ such that $P(A_{N_5,\epsilon_5})=1$, where
\begin{eqnarray*}
A_{n,\epsilon}=\{ \arg \max_{y_2} \,\ f^k(y_2|I_1^k) \in [\underline{b}_1^3 \,\ BR_2(\underline{b}_2^3)+\epsilon] \forall k \geq n \}.
\end{eqnarray*}
Because $\overline{b}_1^3< BR_2(\underline{b}_2^1)$ and because according to Proposition 2, for the given $0 < \epsilon_6 < BR_1(\overline{b}_1^3)-BR_1(BR_2(\underline{b}_2^1))$ there exists an $N_6$ such that $P(B_{N_6,\epsilon_6})=1$, where
\begin{eqnarray*}
B_{n,\epsilon}=\{ \arg \max_{y_1} \,\ f^k(y_1|I_2^k) \in [BR_1(\overline{b}_1^3)-\epsilon \,\ \overline{b}_2^3] \forall k \geq n \}.
\end{eqnarray*}
Let define,
\begin{eqnarray*}
&&[\underline{b}_1^4 \,\ \overline{b}_1^4] := [\underline{b}_1^3 \,\ BR_2(\underline{b}_2^3)+\epsilon_5],\\
&&[\underline{b}_2^4 \,\ \overline{b}_2^4] := [BR_1(\overline{b}_1^3)-\epsilon_6 \,\ \overline{b}_2^3].
\end{eqnarray*}
It is easy to check that $\underline{b}_2^4 > BR_1(BR_2(\underline{b}_2^1))$ and \\ $\overline{b}_1^4 < BR_2(BR_1(\overline{b}_1^1))$.
Similarly, a sequence $N_{n},N_{n+1},...$ can be found such that
\begin{eqnarray*}
\overline{b}_1^{N_n} < \underbrace{BR_2(BR_1(BR_2...BR_1(\overline{b}_1^1)))}_{n times}\end{eqnarray*}
and
\begin{eqnarray*}
\underline{b}_2^{N_n} > \underbrace{BR_1(BR_2(BR_1...BR_2(\underline{b}_2^1)))}_{n times}.
\end{eqnarray*}
Let
\begin{eqnarray*}
\Phi_i^n(.) := \underbrace{BR_i(BR_{-i}(...BR_{i}(.)))}_{n times},
\end{eqnarray*}
because of the uniqueness of the Nash equilibrium, $\Phi_i^n(y_i) \longrightarrow y_i^*$. For the given $\delta>0$ there exists an $\tilde{n}$ such that for $n>\tilde{n}$, $|\Phi_i^n(y_i)-y_i^*|<\delta$; hence for $n>N_1+N_2+...+N_{\tilde{n}}$, $|\overline{b}_1^{n}-y_2^*|<\delta$ and $|\underline{b}_2^{n}-y_1^*|<\delta$.

In summary, by using Proposition 1 and 2 an infinite sequence of sets, $\{B_i,B_i^1,B_i^2,...\}$ can be found such that $B_i \supset B_i^1 \supset B_i^2 \supset ...$, in which \\
\begin{enumerate}
\item $B_i^n=[\underline{b}_i^n \,\ \overline{b}_i^n]$,
\\
\item $\arg\max_{y_2} f^m(y_2|I_1^m) \in B_i^n, \,\  m=N_{n},N_{n+1},...$
\\
\item $ \underline{b}_{i}^{n} \longrightarrow \Phi_j^n(\overline{b}_j)$ and $\overline{b}_{i}^{n} \longrightarrow \Phi_j^n(\underline{b}_j)$
\end{enumerate}

Therefore convergence to the Nash equilibrium follows.

%%%%%%%%%%%%%%%%%%%%%%%%%%%%%%%%%%%%%%%%%%%%%%%%%%%%%%%%%%%%%%%%%%%%%%%%%%%%%%%%
\section{CONCLUSIONS}

This work introduce a two-player competitive game in which players do not know the opponent's utility function and the opponent's action. It is shown that players can learn the Nash equilibrium by engaging in a strategic learning process in which they play a one-shot game repeatedly. Players construct a belief about their opponent's strategy set and play their best response to the Maximium A Posteriori (MAP) of their beliefs. In every stage players observe their own sales and update their beliefs about their opponent's strategy set. It is proven that players' beliefs and actions will converge to the Nash equilibrium. Future work can be done that would investigate the case in which players do not know their opponent's local demand distribution.

\section{Appendix}

\subsection{Proof of Proposition 1}

According to \cite{Z1}-\cite{Z4} the Nash equilibrium is unique. It can be claimed that 2) and 3) cannot be satisfied simultaneously unless $(\underline{b}_2,\overline{b}_1)=(y_1^*,y_2^*)$. Similarly, 1) and 4) cannot be satisfied simultaneously unless $(\overline{b}_2,\underline{b}_1)=(y_1^*,y_2^*)$.
\\
\begin{enumerate}
\item $\underline{b}_1 \geq BR_2(\overline{b}_2),$\\
\item $\overline{b}_1  \leq BR_2(\underline{b}_2),$\\
\item $\underline{b}_2 \geq BR_1(\overline{b}_1),$\\
\item $\overline{b}_2  \leq BR_1(\underline{b}_1).$
\end{enumerate}
Suppose 2) and 3) are satisfied, since best response functions are nonincreasing (Proposition 1 part 2b); therefore,
\begin{eqnarray*}
\overline{b}_1 \leq BR_2(\underline{b}_2) &\leq& BR_2(BR_1(\overline{b}_1)) \\
&\leq& BR_2(BR_1(BR_2(\underline{b}_2))) \\
&\leq& BR_2(BR_1(BR_2(BR_1(\overline{b}_1))))\\
&\leq& \cdots; \,\ (5)
\end{eqnarray*}
the odd-numbered inequalities result from 2) and the even-numbered inequalities result from 3). From Proposition 1 part 4) it is evident that (5) contradicts the uniqueness of the Nash equilibrium unless $(\underline{b}_2,\overline{b}_1)=(y_1^*,y_2^*)$ because the initial assumption was that $y_2^* \in [\underline{b}_1 \,\ \overline{b}_1]$.

Similarly, suppose 1) and 4) are satisfied, then
\begin{eqnarray*}
\underline{b}_1 \geq BR_2(\overline{b}_2) &\geq& BR_2(BR_1(\underline{b}_1))  \\
&\geq& BR_2(BR_1(BR_2(\overline{b}_2)))  \\
&\geq& BR_2(BR_1(BR_2(BR_1(\underline{b}_1))))  \\
&\geq& \cdots;\,\ (6)
\end{eqnarray*}
the odd-numbered inequalities result from 1) and the even-numbered inequalities result from 4).
It is evident that (6) contradicts the uniqueness of the Nash equilibrium unless $(\overline{b}_2,\underline{b}_1)=(y_1^*,y_2^*)$,
because the initial assumption was that $y_2^* \in [\underline{b}_1 \,\ \overline{b}_1]$.

It is assumed, without loss of generality, that $\overline{b}_1 > BR_2(\underline{b}_2)$.
Because the best response functions are nonincreasing (Proposition 1), it follows that
\begin{eqnarray*}
\overline{b}_1 > \overline{\beta} = BR_2(\underline{b}_2) &\geq& BR_2(y_1) (7)\\&=& \arg\max_{y_2} G_2(y_1,y_2) \,\  \forall y_1 \in [\underline{b}_2 \,\ \overline{b}_2].
\end{eqnarray*}
Therefore, $y_2^n \notin U=(\overline{\beta} \,\ \overline{b}_1], \,\ n=1,2,...$. \\
The case $U=[\underline{b}_1 \,\ \underline{\beta})$ and $U=[\underline{b}_1 \,\ \underline{\beta}) \cup (\overline{\beta} \,\ \overline{b}_1]$ can be proved in the same manner.

\addtolength{\textheight}{-3cm}   % This command serves to balance the column lengths

%%%%%%%%%%%%%%%%%%%%%%%%%%%%%%%%%%%%%%%%%%%%%%%%%%%%%%%%%%%%%%%%%%%%%%%%%%%%%%%%

\subsection{Proof of Proposition 3}
It is easy to check that
\begin{eqnarray*}
&&f(y_2|I_1^{n+1})=\frac{f(y_2|I_1^1)f(s_1^1|y_2,y_1^1)...f(s_1^n|y_2,y_1^n)}{\int f(y_2|I_1^1)f(s_1^1|y_2,y_1^1)...f(s_1^n|y_2,y_1^n)dy_2},\\
&&\max_{y_2}log f(y_2|I_1^{n+1})=\max_{y_2}\sum_{m=1}^{n}log f(s_1^m|y_2,y_1^m).
\end{eqnarray*}
Lemma: For every $y_2\neq y_2^n$\\
$E \,\ \{ log f(s_1^n|y_2,y_1^n) \} < E \,\ \{ log f(s_1^n|y_2^n,y_1^n)\}$,\\
where the conditional expectation is taken with respect to $s_1^n$ given $y_2^n$.\\
Proof:
\begin{eqnarray*}
&&E \,\ \{log f(s_1^n|y_2,y_1^n)\}-E \,\ \{log f(s_1^n|y_2^n,y_1^n)\}\\
&&=E \,\ \{ log f(s_1^n|y_2,y_1^n)-\,\ log f(s_1^n|y_2^n,y_1^n) \}\\
&&=E \,\ \{ log \frac{f(s_1^n|y_2,y_1^n)}{f(s_1^n|y_2^n,y_1^n)} \}\\
&&\leq \\
&&log\{E\,\ \frac{f(s_1^n|y_2,y_1^n)}{f(s_1^n|y_2^n,y_1^n)} \}\\
&&=log \{\int_{A} \frac{f(s_1^n|y_2,y_1^n)}{f(s_1^n|y_2^n,y_1^n)}f(s_1^n|y_2^n,y_1^n) ds_1^n \}\\
&&=log \{\int_{A} f(s_1^n|y_2,y_1^n) ds_1^n \}
\end{eqnarray*}
in which $A=\{s_1^n: f(s_1^n|y_2^n,y_1^n)>0\}$. \\
It is evident that $\int_{A} f(s_1^n|y_2,y_1^n) ds_1^n<1$, unless $y_2=y_2^n$; therefore,
\begin{eqnarray*}
E \{log f(s_1^n|y_2,y_1^n)\} < E \{log f(s_1^n|y_2^n,y_1^n)\}
\end{eqnarray*}
and by taking sum over $n$
\begin{eqnarray*}
\frac{1}{N}\sum_{n=1}^N E \{log f(s_1^n|y_2,y_1^n)\} < \frac{1}{N}\sum_{n=1}^N E \{log f(s_1^n|y_2^n,y_1^n)\}.
\end{eqnarray*}
From [8] (page 418)
\begin{eqnarray*}
\frac{1}{N}\sum_{n=1}^N E \{log f(s_1^n|y_2,y_1^n)\}- \frac{1}{N}\sum_{n=1}^N \{log f(s_1^n|y_2,y_1^n)\}\longrightarrow  0,
\end{eqnarray*}
\begin{eqnarray*}
\frac{1}{N}\sum_{n=1}^N E \{log f(s_1^n|y_2^n,y_1^n)\}-\frac{1}{N}\sum_{n=1}^N  \{log f(s_1^n|y_2^n,y_1^n)\}\longrightarrow  0,
\end{eqnarray*}
therefore
\begin{eqnarray*}
\lim \frac{1}{N}\sum_{n=1}^N \{log f(s_1^n|y_2,y_1^n)\} < \lim \frac{1}{N}\sum_{n=1}^N  \{log f(s_1^n|y_2^n,y_1^n)\}.
\end{eqnarray*}
and let $Y_2$ be $Y_2=\{y_2^1,y_2^2,...\}$, then
\begin{eqnarray*}
&\sup_{y_2 \notin Y_2} \lim \frac{1}{N}\sum_{n=1}^N \{log f(s_1^n|y_2,y_1^n)\}& \\
&<& \\
&\sup_{y_2 \in Y_2} \lim \frac{1}{N}\sum_{n=1}^N  \{log f(s_1^n|y_2,y_1^n)\}.&
\end{eqnarray*}

\end{document}